\documentclass[aip,showpacs,reprint,jcp]{revtex4-1}
\usepackage{graphicx}
\pdfoutput=1

\begin{document}

\title{Formation of dodecagonal quasicrystals in two-dimensional 
systems of patchy particles}
\author{Marjolein N.~van der Linden}
\thanks{Present address: Soft Condensed Matter, Debye Institute for 
Nanomaterials Science, Utrecht University, Princetonplein 5, 3584 CC 
Utrecht, The Netherlands}
\affiliation{Physical and Theoretical Chemistry Laboratory, 
  Department of Chemistry, University of Oxford, South Parks Road, 
  Oxford, OX1 3QZ, United Kingdom}
\author{Jonathan P.~K.~Doye}
\thanks{Author for correspondence}
\affiliation{Physical and Theoretical Chemistry Laboratory, 
  Department of Chemistry, University of Oxford, South Parks Road, 
  Oxford, OX1 3QZ, United Kingdom}
\author{Ard A.~Louis}
\affiliation{Rudolf Peierls Centre for Theoretical Physics, 
 University of Oxford, 1 Keble Road, Oxford, OX1 3NP, United Kingdom}

\date{\today}

\begin{abstract}
The behaviour of two-dimensional patchy particles with 5 and 7 
regularly-arranged patches is investigated by computer simulation. 
For higher pressures and wider
patch widths, hexagonal crystals have the lowest enthalpy, whereas at lower 
pressures and for narrower patches, lower-density crystals with five 
nearest neighbours and that are based on the 
($3^2$,4,3,4) tiling of squares and triangles become lower in enthalpy.
Interestingly, in regions of parameter space near
to that where the hexagonal crystals become stable, 
quasicrystalline structures with dodecagonal symmetry 
form on cooling from high temperature. These quasicrystals can be
considered as tilings of squares and triangles, and are probably stabilized
by the large configurational entropy associated with all 
the different possible such tilings. 
The potential for experimentally realizing such
structures using DNA multi-arm motifs are discussed.
\end{abstract}
\maketitle

\section{Introduction}
Since the formation of quasicrystals were first reported in 1984 by
Shechtman {\it et al.} for an Al-Mn alloy,\cite{Shechtman84}
many systems have been to found to exhibit a quasicrystalline phase.
Most of these are binary or ternary metallic alloys (but never a pure metal).
The quest to find quasicrystals beyond alloys has led to an increasing 
number of examples, mostly  in the field of 
soft condensed matter.\cite{Zeng05} 
Examples include dendrimers,\cite{Zeng04,Percec09} 
star polymers,\cite{Hayashida07} micelles\cite{Fischer11}
and binary mixtures of nanoparticles.\cite{Talapin09}
All of these examples, except for one of the micellar systems,\cite{Fischer11}
have dodecagonal symmetry and are often found in regions of 
parameter space close to where crystalline approximants, 
such as the Frank-Kasper $\sigma$ phase, are 
observed.\cite{Chen05,Takano05,Shevchenko06,Matsushita10,Lee10,Zeng11}

A remaining target is to find a colloidal system that can self-assemble into
a quasicrystalline structure in the absence of an external field 
(two-dimensional colloidal quasicrystals can be induced to form using 
quasiperiodic light fields\cite{Mikhael08,Mikhael10,Mikhael11}). 
One approach might be to use a binary or ternary colloidal mixture, but
although complex crystal structures have been reported for binary 
mixtures,\cite{Bartlett92,Eldridge93a,Eldridge93b,Hynninen06b,Hynninen07} 
as yet no quasicrystals have been observed. 
Another approach might be to use colloids with anisotropic ``patchy'' 
interactions,\cite{Glotzer07} where the positions of the patches could be used 
to control the preferred local geometry and hence influence the global 
structure formed. 
Indeed, much progress has been made in developing methods
to synthesize such types of colloidal 
particles.\cite{Manoharan03,Cho05,Cho07,Pawar08,Perro09,Kraft09b,Perro10,Jiang10,Kraft11,Duguet11} 
Furthermore, the first experiments on the novel structures that 
such patchy interactions can enable the systems to adopt are beginning to 
appear,\cite{Chen11b} as well as being systematically explored through
computer simulations.\cite{Zhang06,Doye07,Noya07b,Noya10,Romano09,Romano11b,Doppelbauer10,Bianchi11,Romano11,Antlanger11}

Quasicrystals have also been found to form for a variety of model 
potentials in computer simulations. Interestingly, these are not restricted
to mixtures,\cite{Widom87,Leung89} but can also occur for one-component 
systems.\cite{Jagla98,Skibinsky99,Dzugutov93,Quandt99,Keys07,Engel07,Engel10,Iacovella11,HajiAkbari09,HajiAkbari11b,Johnston10b,Johnston11} 
Examples include isotropic pair potentials with both a maximum and a 
minimum in the potential,\cite{Dzugutov93,Quandt99,Keys07,Engel07,Engel10} micellar 
models,\cite{Iacovella11} hard tetrahedra\cite{HajiAkbari09} and triangular 
bipyramids,\cite{HajiAkbari11b} 
and water\cite{Johnston10b} and silicon\cite{Johnston11} bilayers.

Here, we wish to examine in detail the behaviour of two-dimensional patchy 
particles with 5 and 7 patches as possible quasicrystal-forming systems. 
The reason we choose this system is that in a recent study 
of two-dimensional disks with regularly arranged patches, intriguing behaviour
was seen for the 5-patch system.\cite{Doye07,Doppelbauer10}
Hard disks naturally form a hexagonal crystal, and six patches reinforce 
this tendency. 
For four patches, there is a competition between
a square crystal, which is energetically stabilized by the patchy 
interactions, and a hexagonal crystal, which is stabilized by its higher 
density. 
However, the situation for particles with 5 regularly arranged 
patches is more complex, since the 5-fold symmetry of the particles 
is incompatible with crystalline order --- there is no crystal in which all
the patches can point directly at those on neighbouring particles. 

\begin{figure}
\includegraphics[width=8.4cm]{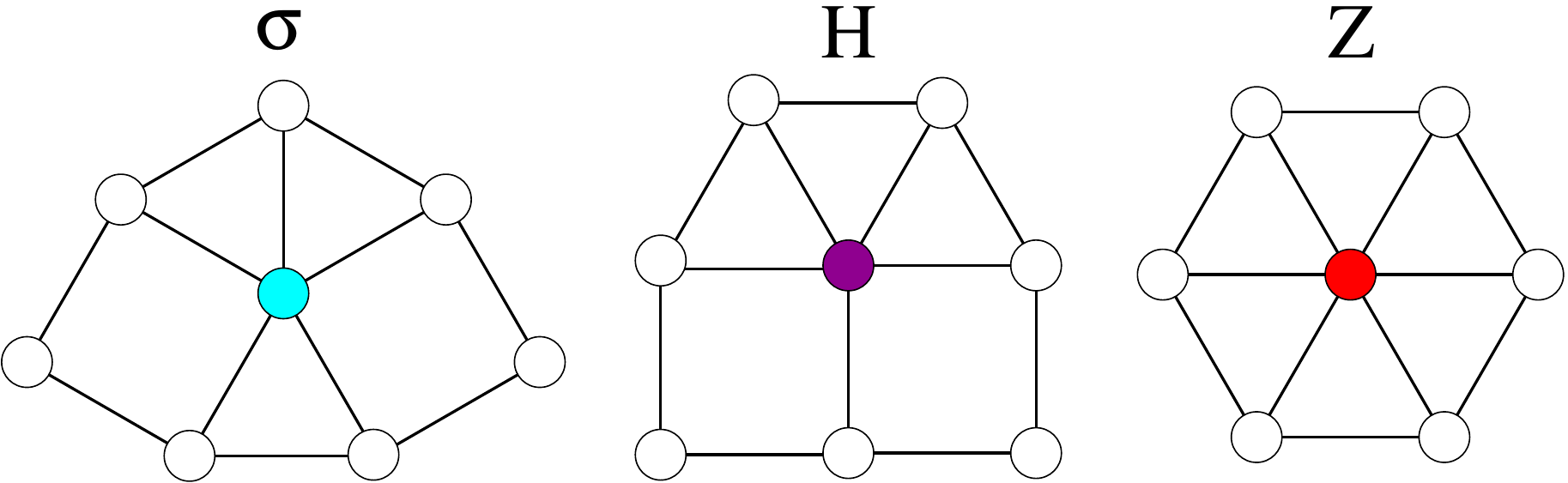}
\caption{\label{fig:local} The three local environments that are mainly
present in the simulated structures. 
}
\end{figure}

In Ref.\ \onlinecite{Doye07} we found that on
cooling (at low pressure), crystallization was not observed, because of the 
frustration introduced by the geometry of the patches.
The configurations generated did though show certain common local motifs
(those shown in Fig.\ \ref{fig:local}) and could be considered
as tilings of squares and triangles. 
Although there was no overall crystalline order, there was some 
evidence suggestive of longer-range orientational order; however,
because of the relatively small systems sizes considered this was not pursued 
further at the time. Given the known tendency of random 
square-triangle tilings to form dodecagonal quasicrystals,\cite{Oxborrow93} 
we now investigate 
in detail the suggestion made in Ref.\ \onlinecite{Doye07} that this system 
may form quasicrystals.

\section{Methods}
\subsection{Potential}
The model consists of spherical particles patterned with attractive patches. They are described by a modified Lennard-Jones potential,
in which the repulsive part of the potential is isotropic, but the attractive part is anisotropic and depends on the alignment of patches
on interacting particles. Specifically, the potential is described by
\begin{equation}
\label{eq:potential}
V_{ij}({\mathbf r_{ij}},{\mathbf \Omega_i},{\mathbf \Omega_j})=\left\{
    \begin{array}{ll}
       V_{\rm LJ}(r_{ij}) & r<\sigma_{\rm LJ} \\
       V_{\rm LJ}(r_{ij})
       V_{\rm ang}({\mathbf {\hat r}_{ij}},{\mathbf \Omega_i},{\mathbf \Omega_j})
                       & r\ge \sigma_{\rm LJ}, \end{array} \right.
\end{equation}
where $V_{\rm LJ}$, the Lennard-Jones potential, is given by
\begin{equation}\label{eqn:LJ} V_{\rm LJ}(r) = 4\epsilon\left[ \left( \frac{\sigma_{\rm LJ}}{r}
    \right)^{12} - \left( \frac{\sigma_{\rm LJ}}{r} \right)^{6} \right].
\end{equation}
The minimum of this potential is at $r_{\rm eq}=2^{1/6}\sigma_{\rm LJ}$.
$V_{\rm ang}$ is an angular modulation factor that depends on the orientations of the patches on the two interacting particles with respect to the 
interparticle vector. Specifically,
\begin{equation}
V_{\rm ang}({\mathbf {\hat r}_{ij}},{\mathbf \Omega_i},{\mathbf \Omega_j})=
G_{ij}({\mathbf {\hat r}_{ij}},{\mathbf \Omega_i})
G_{ji}({\mathbf {\hat r}_{ji}},{\mathbf \Omega_j}),
\end{equation}
where
\begin{equation}
\label{eqn:AngMod}
G_{ij}({\mathbf {\hat r}_{ij}},{\mathbf \Omega_i})=
\exp\left(-\frac{\theta_{k_{\rm min}ij}^2}{2\sigma_{\rm pw}^2}\right),
\end{equation}
$\sigma_{\rm pw}$ is a measure of the angular width of the patches, 
$\theta_{kij}$ is the angle between patch vector $k$ on particle $i$
and the interparticle vector $\mathbf r_{ij}$, and $k_{\rm min}$ is the patch that minimizes the magnitude of this angle. Hence,
only the patches on each particle that are closest to the interparticle vector interact with each other, and $V_{\rm ang}=1$ if the patches point directly at 
each other.
One feature of this potential is that as $\sigma_{\rm pw}\rightarrow\infty$ the isotropic Lennard-Jones potential is recovered.
For computational efficiency the potential is truncated and shifted at $r=3\,\sigma_{\rm LJ}$, and the crossover distance in Eq.\ \ref{eq:potential}
is adjusted so that it still occurs when the potential is zero.

As well as studying crystalization in two 
dimensions,\cite{Doye07,Doppelbauer10,Antlanger11}
this patchy particle model has also been previously used to study 
crystallization in three dimensions,\cite{Doye07,Noya07b,Noya10} and 
the self-assembly of monodisperse shells.\cite{Wilber07,Wilber09,Williamson11}

\subsection{Simulation}
Systems of particles were simulated using standard Metropolis Monte Carlo (MC)
in the NPT ensemble. Periodic boundary conditions were applied, and 
the simulation box was constrained to be square. 
The MC moves were single-particle rotational and translational moves, 
as well as ``volume'' moves in which the area of the box was scaled, the latter
allowing a constant pressure ensemble to be simulated. 
The number of particles $N$ was always 2500.

\subsection{Structural analysis}

We employ a number of approaches to analyse the structures that the systems of 
patchy particles adopt. In the condensed state, we find that the 5- and 7-patch 
particles virtually always adopt one of three local environments. 
These are illustrated in Fig.\ \ref{fig:local}. 
There are two possible five-coordinate environments, 
which are based upon two different ways of 
locally packing squares and triangles. We refer to them as the $\sigma$ and $H$ 
environments by analogy to the Frank-Kasper phases of these names.\cite{FrankK58,FrankK59} 
These $\sigma$ and $H$ Frank Kasper phases are crystalline approximants to 
dodecagonal quasicrystals and can be considered as square-triangle tilings in 
two of their three dimensions.\cite{Shoemaker} 
The tilings containing only these local environments are more formally
denoted by $(3^2,4,3,4)$ ($\sigma$) and $(3^3,4^2)$ ($H$),
where this nomenclature refers to the sequence of polygons around a 
vertex.\cite{Grunbaum}
The third local environment is the six-coordinate hexagonal environment, 
which represents the densest local packing, and will be denoted $Z$, again
by analogy to a Frank-Kasper phase. 

To identify these environments we employ a common neighbour analysis. 
Neighbours are defined as all particles within a certain cutoff radius $r_c$ 
from a given particle. Then for each pair of neighbours the number of neighbours
common to both is determined. Each local environment is characterized by a 
unique signature in terms of the number of neighbours with which the central
particle shares a certain number of neighbours. For example, 
a particle in a $\sigma$ local environment has a total of five neighbours, 
with one of which it has two neighbours in common and with four of which it has 
one neighbour in common. Hence, this environment is denoted by the common
neighbour signature \{21111\}. In a similar way, the signatures for the 
$H$ and $Z$ local environments are \{22110\} and \{222222\}, respectively. 
Particles that are not in any of these three local environments are 
labelled `undefined' ($U$). The cutoff radius 
($r_{c}\approx1.38\,\sigma_{\rm LJ}$) was chosen 
such that the fraction of $U$-particles was minimized.

To probe the global order of the configurations generated, the associated
diffraction patterns were calculated. Quasicrystalline configurations are 
characterized by diffraction patterns with non-crystallographic rotational 
symmetries. We calculated the diffraction patterns by evaluating the
real part of the interference function:
\begin{equation}
\label{eq:IntFunc}
S(\mathbf{q})=\frac{1}{N}\sum_{i=0}^{N-1}\sum_{j=0}^{N-1}\exp\left[2\pi\textrm{i}\mathbf{q}\cdot(\mathbf{r}_{i}-\mathbf{r}_{j})\right],
\end{equation}
where $\mathbf{q}$
is the wavevector and 
$\mathbf{r}_{i}$
the position of particle $i$.
Unless stated otherwise, the resolution we used for the plots of the 
diffraction patterns is $\Delta q_{x}=\Delta q_{y}=0.15\,\sigma_{\rm LJ}^{-1}$.

When mapping out the behaviour of our system as a function of the parameter 
space it will be useful to to measure the degree of twelvefold symmetry of 
the diffraction pattern. To achieve this, we take a Fourier transform around a
ring of the interference pattern at a $q$ value corresponding to the first
diffraction peaks. Specifically, we evaluate 
\begin{equation}
\label{eq:FT}
F(\nu)=\sum_{i=0}^{n-1}S(q_{\rm 1st},\phi_{i})\exp\left(-2\pi\textrm{i}\nu \phi_{i}\right),
\end{equation}
where the sum is over the $n$ sampled points of $S(q_{\rm 1st},\phi)$. 
A twelvefold symmetric diffraction pattern will have high values at 
$\nu=12$ and multiples thereof. Similarly, a sixfold symmetric
diffraction pattern, as would be expected for a hexagonal crystal, 
will have high values at $\nu=6$ and multiples thereof (including $\nu=12$).
Therefore, we used $|F(\nu=12)|- |F(\nu=6)|$ as a measure of the twelvefold
character of the diffraction pattern.
We note that the peaks at multiples of the lowest frequency will have a
lower amplitude than the main lowest frequency peak, where this decay in 
amplitude is stronger if the diffraction peaks are more diffuse.

\begin{figure}
\includegraphics[width=8.4cm]{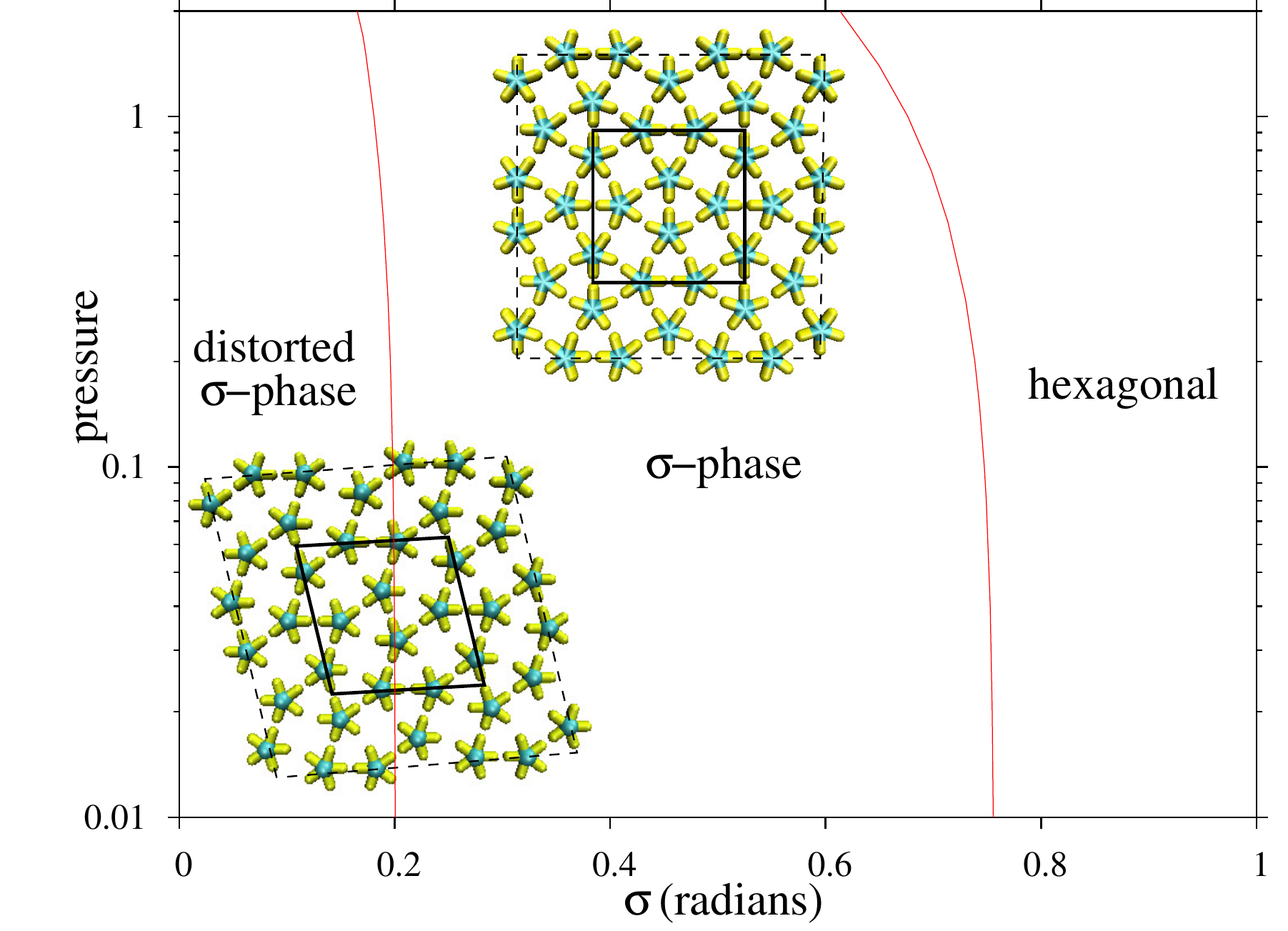}
\caption{\label{fig:xtals}
Zero-temperature phase diagram showing the dependence of the lowest-enthalpy
structure on pressure and patch width. 
}
\end{figure}

\section{Results}
We will first consider the 5-patch particles in detail. The behaviour of the 
7-patch particles is fairly similar and so we will consider this system more 
briefly in Section \ref{sect:7patch}.
\subsection{5-patch particles}
\label{sect:5patch}
\subsubsection{Low-enthalpy structures}
\label{sect:xtals}
The most stable phase at zero temperature is simply that with the lowest
enthalpy. Therefore, we minimized the enthalpy for a variety of candidate
crystal structures under different conditions and potential parameters in 
order to obtain Figure \ref{fig:xtals}, which shows how the lowest enthalpy 
crystal structure depends on pressure and patch width. 
This figure is slightly different from that which appeared in 
Ref.\ \onlinecite{Doye07}, because a slightly lower energy crystal 
at low $\sigma_{\rm pw}$ was subsequently identified in 
Ref.\ \onlinecite{Doppelbauer10}

At sufficiently high pressure the most stable phase will be the crystal with 
highest density, which in this system is the hexagonal crystal. 
At sufficiently low pressure, the crystal structure with 
the lowest energy will be most stable, and will be the one which maximizes
the patch-patch interactions. 

\begin{figure}
\includegraphics[width=8.4cm]{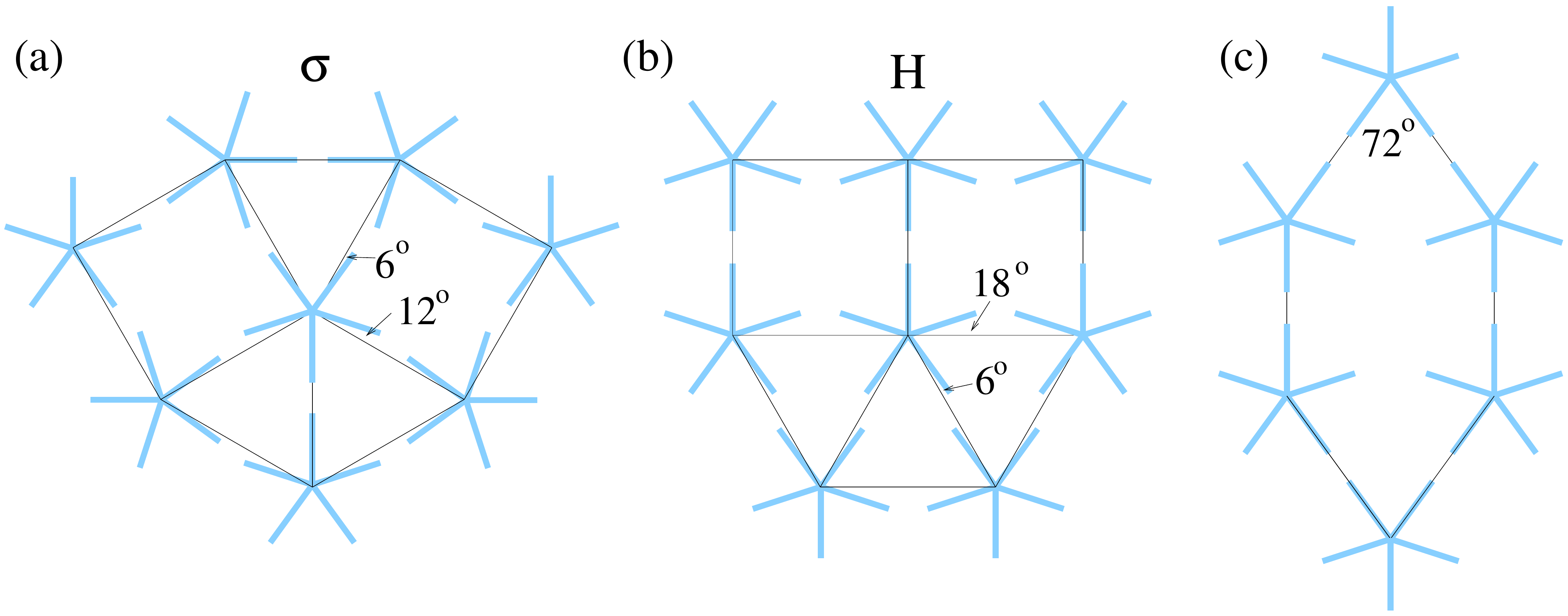}
\caption{\label{fig:angles}
Deviations of the patch vectors from the interparticle vectors in the
(a) $\sigma$- and (b) $H$-type environments. 
(c) At low values of the patch width it is favourable for these structures to
distort so that three of the patches point directly at the neighbouring particles.
The resulting structures can be viewed as being made up of the depicted hexagon.
}
\end{figure}

At intermediate values of the patch width, 
the low energy crystals are those based on the two five-coordinate local 
environments depicted in Fig.\ \ref{fig:local}. Although the patches are
not able to point directly at those on neighbouring particles in these motifs,
the loss in energy is not prohibitive because the angular deviations
are relatively small (Fig. \ref{fig:angles}). 
As the average angular deviation is smaller in the $\sigma$ local environment,
the $\sigma$ crystal is slightly lower in energy than the $H$ crystal.

As the patches becomes narrower, the energetic penalty associated with
the non-perfect alignment of the patches with the interparticle vectors
increases, until a point is reached where it becomes favourable for the local 
environments to distort so that the patches point directly at three of the five 
neighbours. The resulting packings can be considered to be made of the
irregular hexagon shown in Fig.\ \ref{fig:angles}(c).
For the $H$ crystal this distortion does not lead to any change in its space 
group, namely it remains as $cmm$. However, for the $\sigma$ crystal there is a 
symmetry breaking --- 
there are two equivalent ways of dividing up the structure into these hexagons
--- and the space group changes from $p4g$ to $p2$.
The resulting crystals are virtually degenerate, with the distorted $\sigma$ 
structure just lower in energy.\cite{Doppelbauer10}

As the patch width is increased away from the intermediate values at
which the $\sigma$ phase is stable, the orientational dependence of the 
potential becomes weaker, and there comes a point at which the hexagonal 
crystal becomes lowest in energy because of its greater number of neighbours.

In the $\sigma$-phase the ratio of triangles to squares is 2. 
There are also many larger-unit cell crystal structures with a mixture of 
$\sigma$ and $Z$ environments that have an increasing ratio of triangles to 
squares.  We thought that these more complex crystals might be most stable 
near to the $\sigma$/hexagonal boundary in Fig.\ \ref{fig:xtals}.
However, we never found an instance where one of these structures had 
the lowest enthalpy.

\subsubsection{Annealing}
We do not expect quasicrystalline configurations to be lowest in enthalpy, 
both because the closest quasicrystalline approximants (the crystals 
involving both $\sigma$ and $Z$ local environments) are never lowest
in enthalpy (Sect. \ref{sect:xtals}) and because the disorder associated with
quasicrystals is likely to lead them to have a somewhat higher enthalpy than
the relevant approximant.
However, it may be that a quasicrystal is more kinetically accessible than possible crystals on cooling, or is even thermodynamically stable for a particular 
temperature range due to its greater entropy. Therefore, to search for
quasicrystalline behaviour we performed a series of cooling runs for a grid
of pressure and patch width values. In these runs the temperature was
decreased linearly from 0.5$\,\epsilon k^{-1}$ to zero over 50\,000 MC cycles. 
Some of the resulting configurations
did show twelve-fold diffraction patterns characteristic of a dodecagonal 
quasicrystal. However, the peaks were very diffuse. Therefore, we subsequently
annealed all the final configurations at a temperature of $0.15\,\epsilon k^{-1}$
for $10^6$ MC cycles.
This temperature was chosen so that the mobility of the atoms was 
large enough to allow significant ordering during annealing. 
For the most part, this temperature was also below that 
at which the quasicrystals formed, the exceptions occurring near to 
the hexagonal/$\sigma$ boundary in Fig.\ \ref{fig:xtals}, where a hexagonal 
phase formed first on cooling before transforming into the quasicrystalline 
phase at lower temperature. 

\begin{figure}
\includegraphics[width=8.4cm]{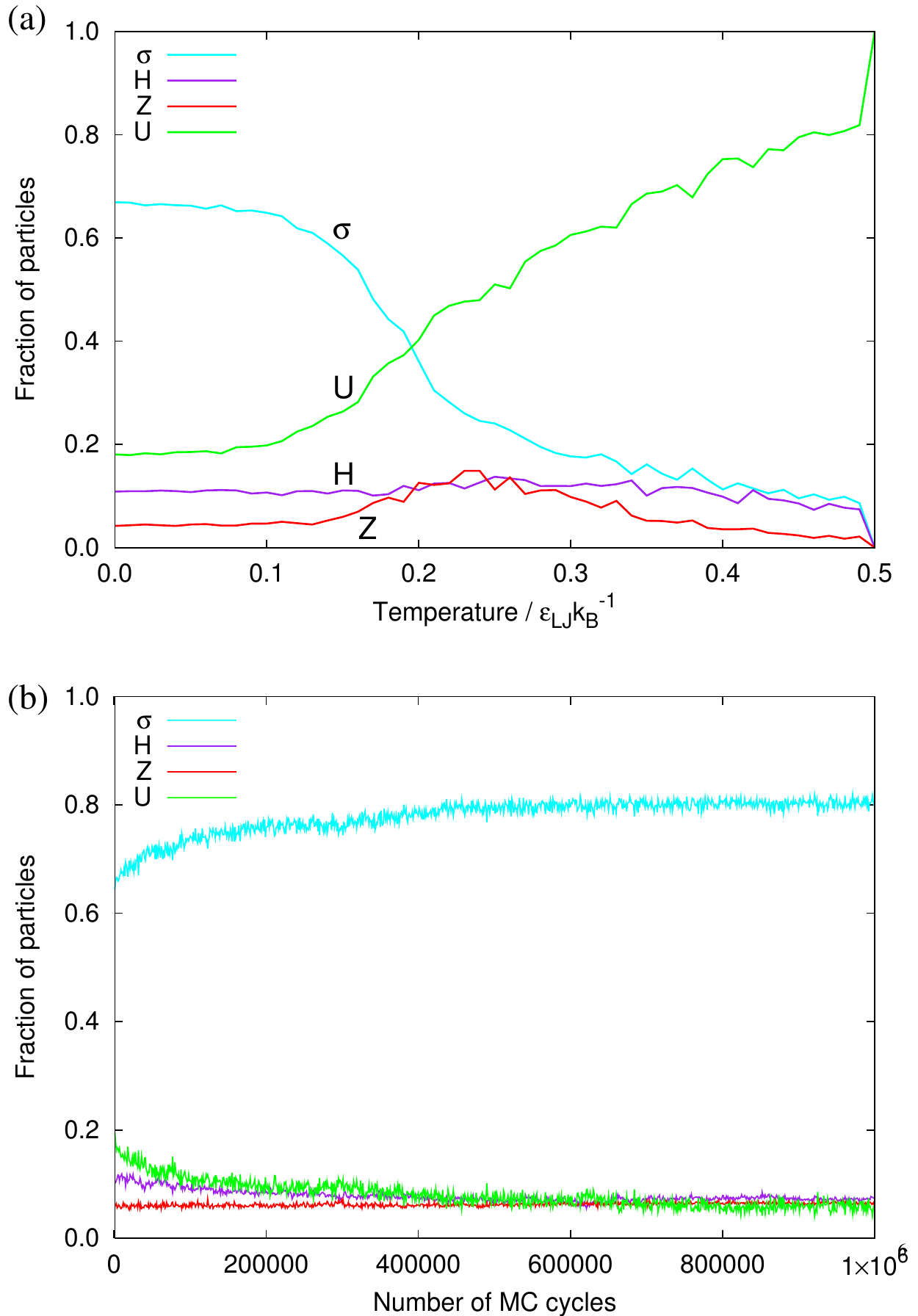}
\caption{\label{fig:5patch_evolution} Evolution of local structural 
environments during (a) cooling and (b) annealing at $T=0.15\epsilon k^{-1}$.
$p=0.7\,\epsilon\sigma_{\rm LJ}^{-2}$, $\sigma_{\rm pw}=0.49$. 
}
\end{figure}

Figure \ref{fig:5patch_evolution} shows the evolution of the number 
of particles with different local environments for the cooling and annealing
runs for a state point that led to a dodecagonal quasicrystal.
On cooling, the number of $\sigma$ environments increases rapidly between 
$0.2$ and $0.1\,\epsilon k^{-1}$ and is due to the formation of the
quasicrystal. Prior to this there is a transient increase
in the number of hexagonal particles. 
Closer to the hexagonal/$\sigma$ boundary in Fig.\ \ref{fig:xtals}, 
this transient increase is more pronounced because of the increased 
stability of the hexagonal phase. Even though the hexagonal crystal is not the 
lowest enthalpy structure under these conditions, it is thermally stabilized by its large orientational entropy.

\begin{figure*}
\includegraphics[width=17cm]{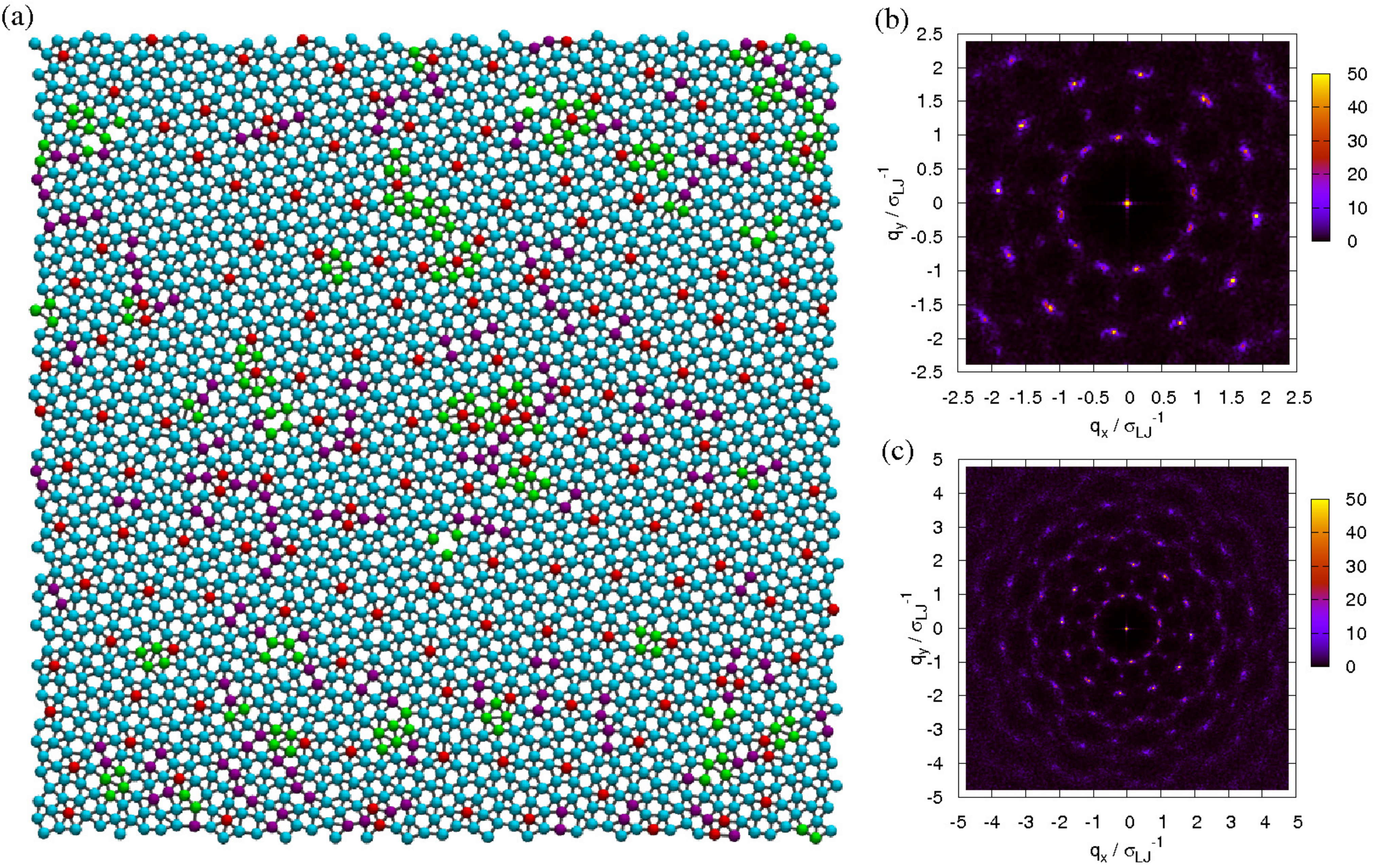}
\caption{\label{fig:5patch_QC} (a) Configuration after annealing and (b) and (c)
associated diffraction patterns (for two different ranges of $\mathbf{q}$) 
at $p=0.7\,\epsilon\sigma_{\rm LJ}^{-2}$,  $\sigma_{\rm pw}=0.49$. 
}
\end{figure*}

It is noteworthy that the number of hexagonal particles does not decrease
to zero on quasicrystal formation. This is not due to incomplete ordering but
is one of the features of the quasicrystalline structures. At the end of the
cooling run, there is still a significant fraction of particles whose local 
structure cannot be assigned because of the disorder within the configurations. 

On annealing the number of $\sigma$ environments gradually increases at the 
expense of unidentified and $H$ environments (but not hexagonal) (Fig.\ \ref{fig:5patch_evolution}(b)). 
The resulting configuration is shown in Fig.\ \ref{fig:5patch_QC} along
with the associated diffraction pattern. The diffraction pattern shows clear
twelvefold symmetry and is very similar to that for a ``perfect'' 
dodecagonal quasicrystal produced by the ``extended Schlottmann'' inflation 
rules\cite{Stampfli88,Hermisson97,Zeng06} (Supplementary Fig.\ 1), 
albeit with less sharp peaks. The twelvefold 
pattern also implies that the orientational order is coherent across the 
whole box; it could be said to be a ``single quasicrystal''. 

\begin{figure}
\includegraphics[width=8.4cm]{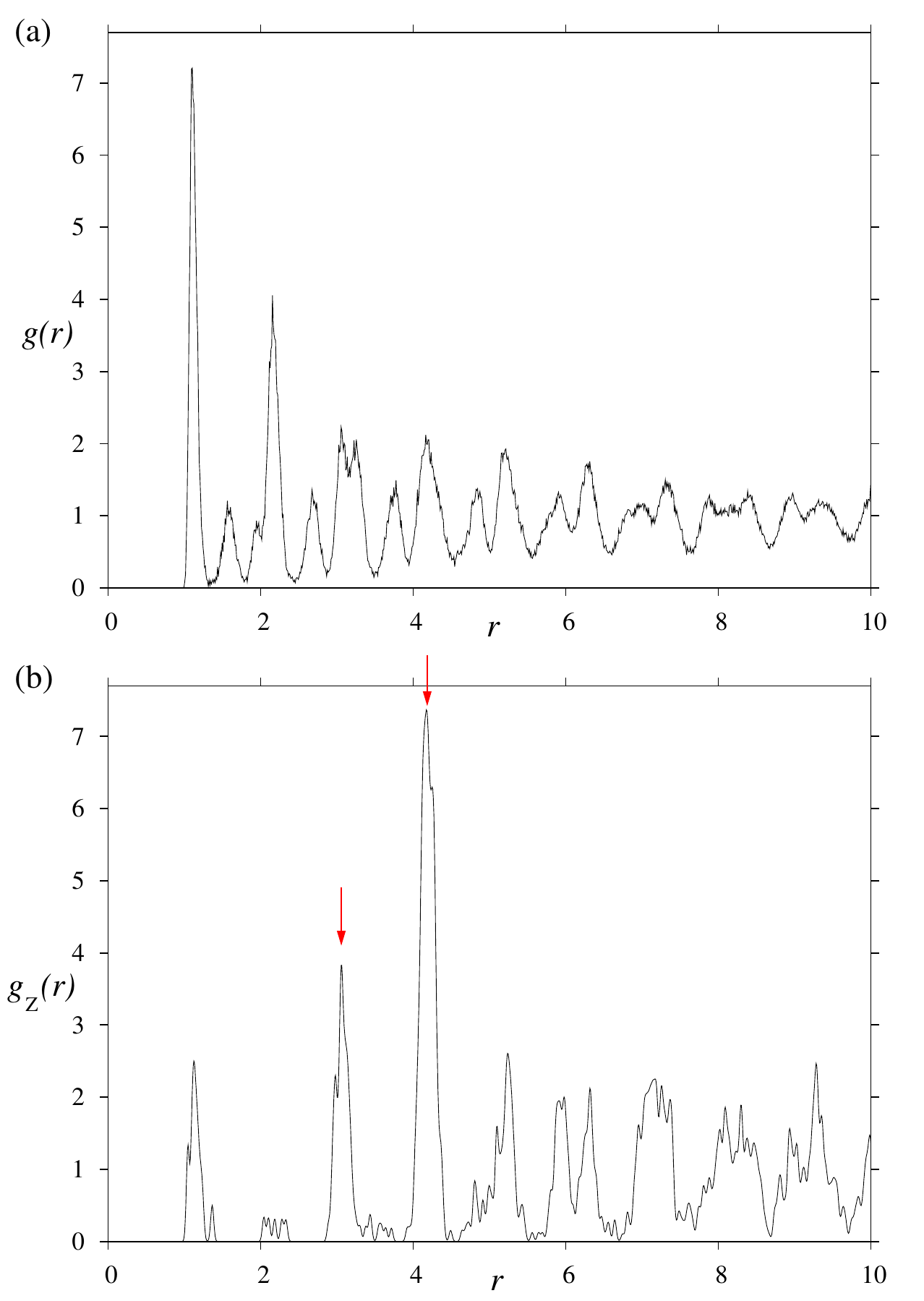}
\caption{\label{fig:rdf} Radial distribution functions 
for (a) all the particles and (b) the $Z$ particles in the quasicrystalline 
configuration obtained at $p=0.7\,\epsilon\sigma_{\rm LJ}^{-2}$,  $\sigma_{\rm pw}=0.49$. 
In (b) the peaks associated with interpenetrating 
(Fig.\ \ref{fig:dodecagon_motifs}(e)) 
and edge-sharing 
(Fig.\ \ref{fig:dodecagon_motifs}(c)) dodecagons 
are marked with arrows.}
\end{figure}

\begin{figure}
\includegraphics[width=8.4cm]{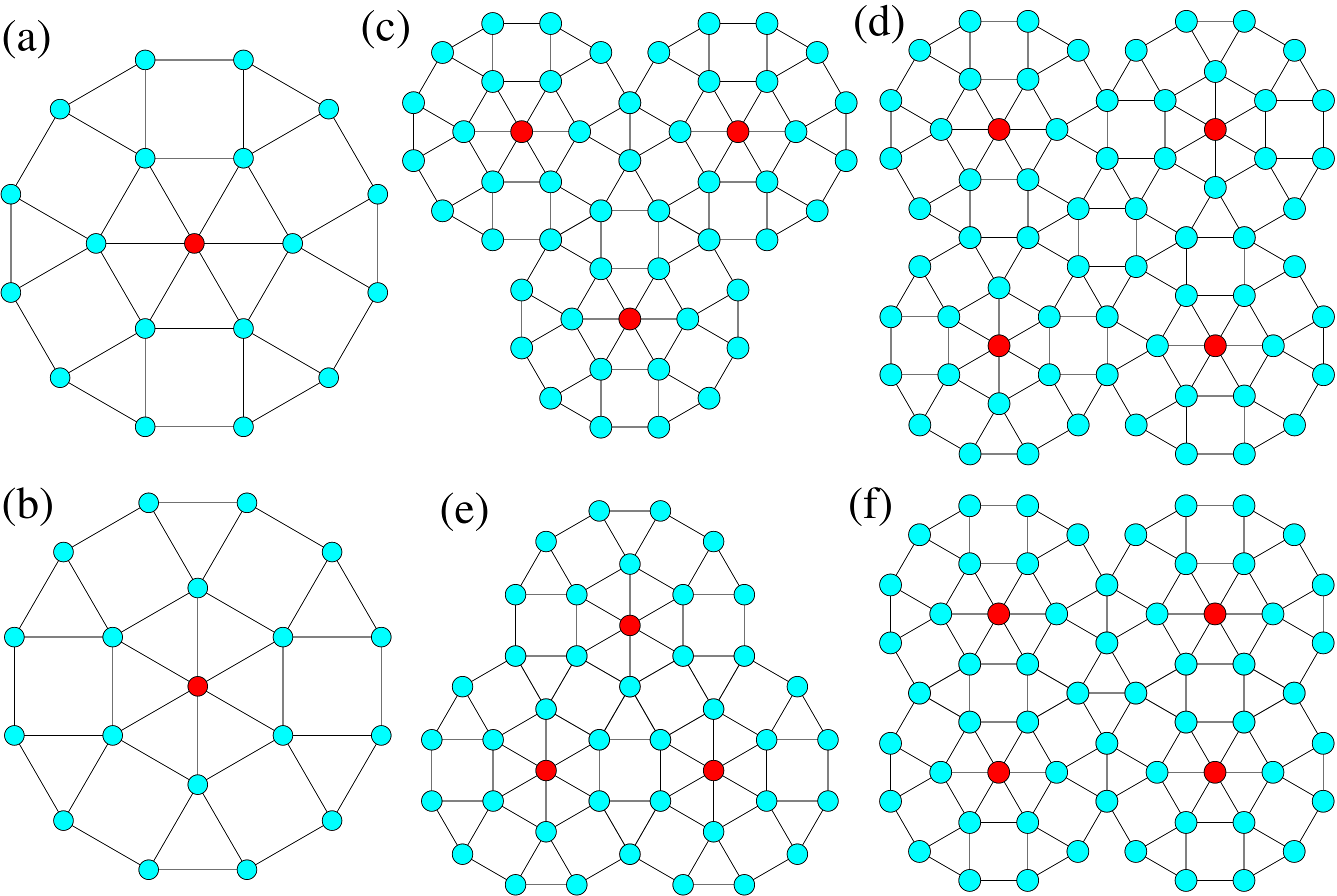}
\caption{\label{fig:dodecagon_motifs}
(a) and (b) Dodecagonal motifs, and (c)--(f) four ways of locally packing these 
motifs.
}
\end{figure}

The configuration is a square-triangle tiling in which
the twelve possible orientations of the bond vectors are equally likely --- hence, the twelvefold symmetry in the diffraction pattern. 
The radial distribution function shows clear peaks out to quite long range
(Fig.\ \ref{fig:rdf}(a)) because of the square-triangle order.
Close inspection of the configuration shows 
that the hexagonal atoms for the most part do not cluster together but are 
instead isolated from each other
and are usually at the centre of the dodecagonal motifs depicted in Fig.\ 
\ref{fig:dodecagon_motifs}(a) and (b). 
These dodecagonal motifs can join together in two ways. 
They can share an edge as in Fig.\ \ref{fig:dodecagon_motifs}(c) 
(their centres are separated by 
$(2+\sqrt 3) r_{\rm eq}=4.189\,\sigma_{\rm LJ}$) 
or interpenetrate as in Fig.\ \ref{fig:dodecagon_motifs}(e) 
(separation $(1+\sqrt 3) r_{\rm eq}=3.067\,\sigma_{\rm LJ}$). 
These two distances are very apparent from the radial distribution function
for particles in hexagonal environments (Fig.\ \ref{fig:rdf}(b)) with a clear 
preference for edge-sharing dodecagons.

The four basic ways of locally packing the dodecagons 
that are possible using interpenetrating and edge-sharing dodecagons
are illustrated in Fig.\ \ref{fig:dodecagon_motifs}(c)--(f),
two of which are triangular, one of which is square and one of which
is rectangular. These motifs can be used to construct a whole variety of 
crystal structures with large unit cells (and varying ratios of squares to
triangles), although as mentioned in Sect.\ \ref{sect:xtals} 
we did not find an instance where these crystals had the lowest enthalpy. 
The model quasicrystal that we created using the extended Schlottmann 
inflation rules 
(Supplementary Fig.\ 1) 
can also be analysed 
in terms of the edge-sharing dodecagon motifs of Fig.\ 
\ref{fig:dodecagon_motifs}(c) and (d).

When the configuration in Fig.\ \ref{fig:5patch_QC}(a) is examined all four
of these motifs can be found, with the triangle of edge-sharing dodecagons
(Fig.\ \ref{fig:dodecagon_motifs}(c)) most common. When examining how these
triangular, square and rectangular elements begin to tile the plane, we find 
configurations like those in Fig.\ \ref{fig:local}, but now, of course, 
on a longer length scale. Furthermore, this tiling seems to be random and
does not completely tile the plane, because of a significant fraction of
defects, both in terms of $U$ particles with an ``unidentified'' coordination
shell and strings of $H$ particles. It is noteworthy that $\sigma$ and $H$ crystals can form a coherent interface along the \{11\} directions of the $\sigma$ crystals, and the latter defects seem to often occur when the $\sigma$-like order
in two adjacent regions is not fully in registry, and needs a line of $H$ atoms
to bridge the regions. 

Another source of disorder is the two possible 
orientations of the central hexagon in the dodecagonal motif, as
illustrated in Fig.\ \ref{fig:dodecagon_motifs}(a) and (b). 
The two forms can be transformed into each other by a rotation of the 
hexagon by $\pi/6$. 
For the edge-sharing dodecagons in Fig.\ \ref{fig:dodecagon_motifs} the 
orientations of the hexagons have been chosen so that all the particles 
on the edge of the dodecagons have the more favourable $\sigma$ environments. 
However, some relative orientations of the internal hexagons lead to $H$ 
environments. 
For example, there are two $H$ environments at the shared edge if both hexagons 
have bonds parallel to the shared edge, and a few examples of such $H$ 
``dimers'' can be found in Fig.\ \ref{fig:5patch_QC}(a). By contrast, for the 
interpenetrating dodecagons (Fig.\ \ref{fig:dodecagon_motifs}(d) and (e)), 
if the hexagon at the centre of one of the dodecagons is rotated, the 
dodecagonal character of the other centres is lost, although the packing
is still a square-triangle tiling.

\begin{figure*}
\includegraphics[width=14cm]{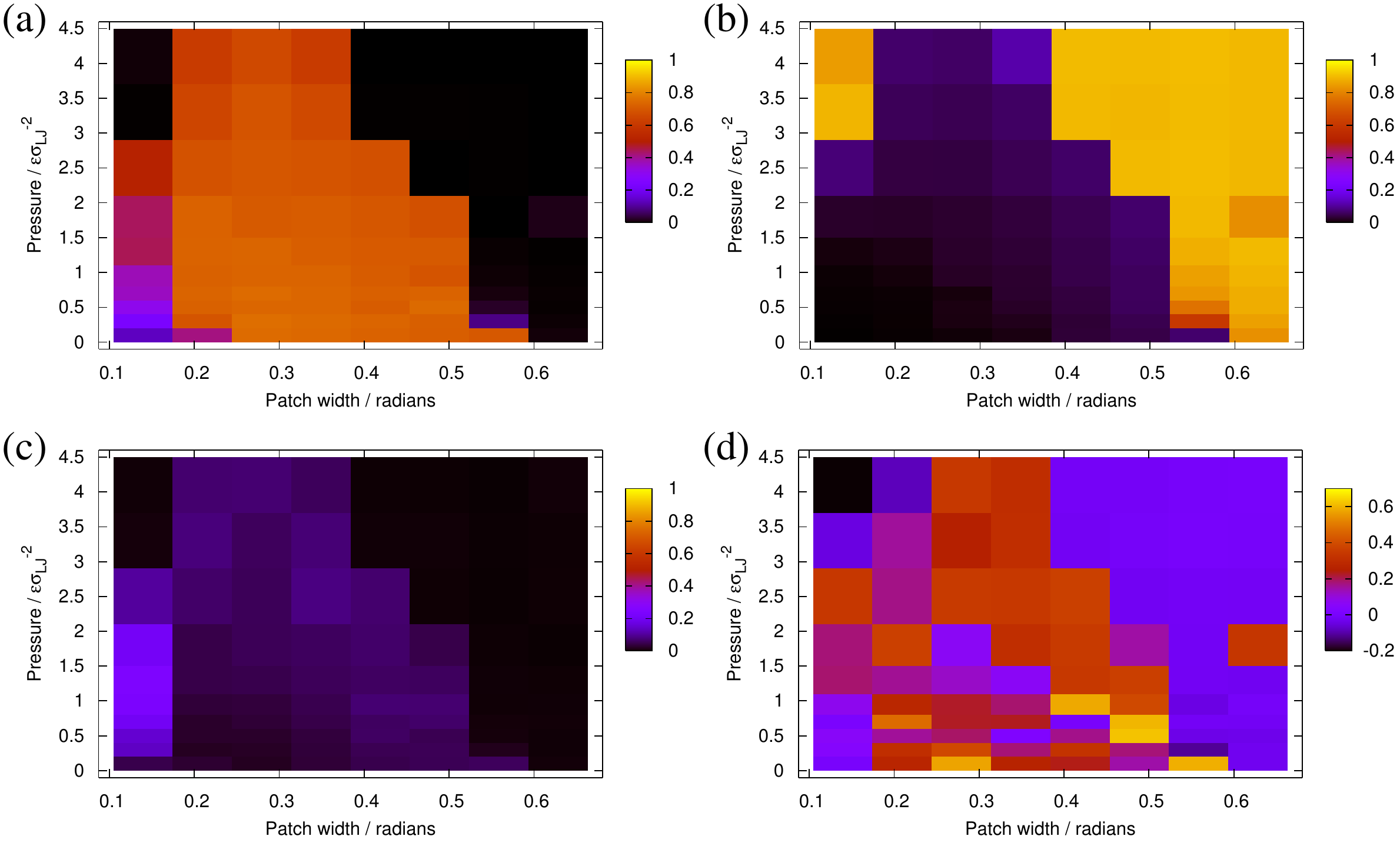}
\caption{\label{fig:5patch_map} Dependence on pressure and patch width 
of the fraction of particles in (a) $\sigma$, (b) $Z$ and (c) $H$ environments, 
and (d) $|F(\nu=12)|- |F(\nu=6)|$, a measure of
the twelvefold character of the diffraction pattern, for the final 
configuration after annealing.
}
\end{figure*}

\begin{figure*}
\includegraphics[width=16cm]{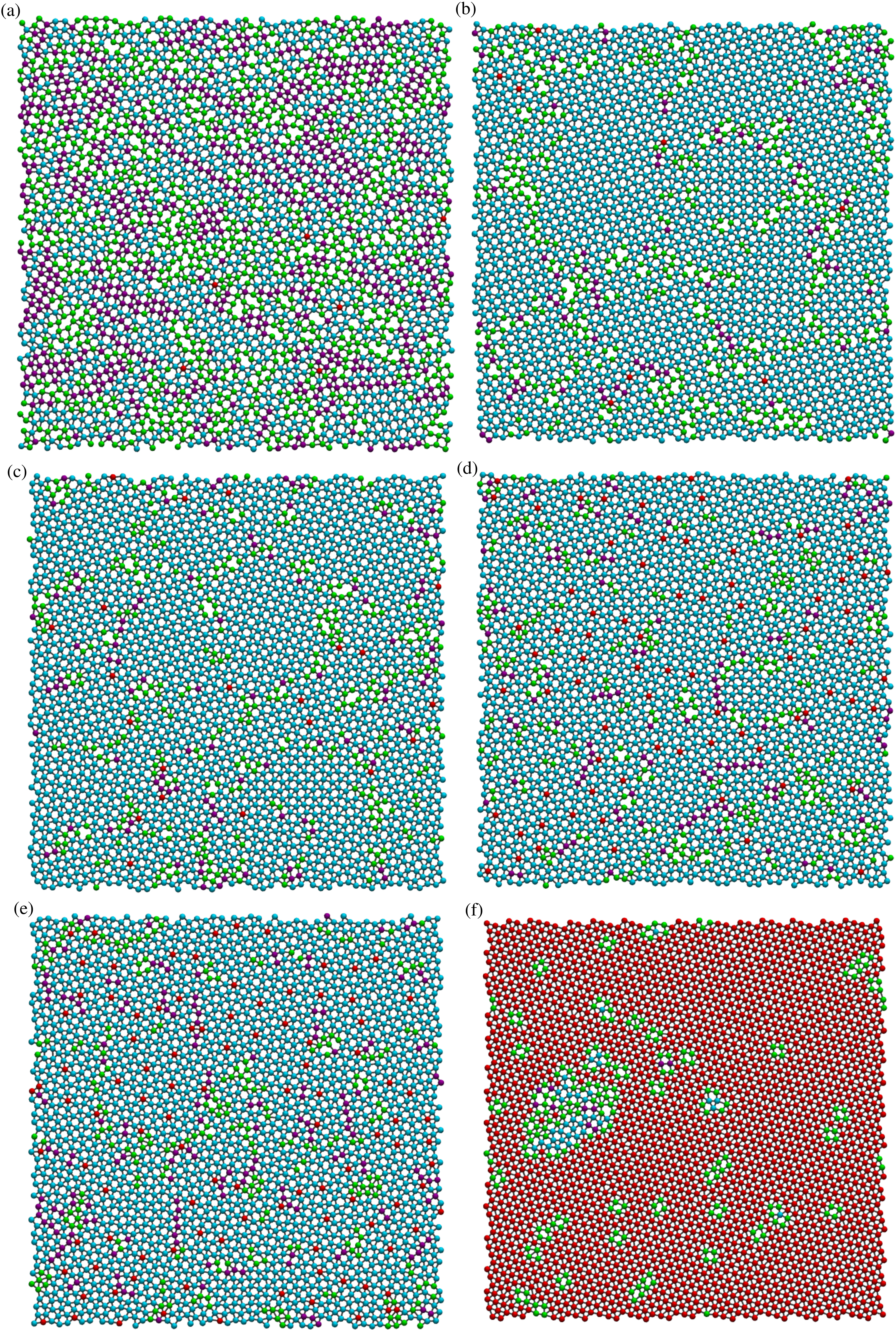}
\caption{\label{fig:5patch_configs} Dependence of the final configuration after 
annealing on patch width (a) $\sigma_{\rm pw}=0.14$, (b) $\sigma_{\rm pw}=0.21$, (c) $\sigma_{\rm pw}=0.28$, (d) $\sigma_{\rm pw}=0.35$, (e) $\sigma_{\rm pw}=0.42$ and (f) $\sigma_{\rm pw}=0.56$. 
All are at $p=0.7\,\epsilon\sigma_{\rm LJ}^{-2}$.  }
\end{figure*}

Another way to characterize the structures that we observe is in terms of
the ratio of triangles to squares in the tiling, and the related ratio
of the number of five- and six-coordinate atoms. For example, we find the 
ratio of triangles to squares for the configuration in 
Fig.\ \ref{fig:5patch_QC}(a) to be 2.303, 
which compares with a value of 2 for the $\sigma$ crystal 
and $4/\sqrt{3}=2.309$ for an ``extended 
Schlottmann'' quasicrystal.\cite{OKeeffe10} 
Similarly, the ratio of 5- to 6-coordinate environments is 13.194 
for the 
configuration in Fig.\ \ref{fig:5patch_QC}(a),
which is again similar to the value for an ``extended Schlottmann'' quasicrystal, namely $(24+14\sqrt3)/(2+\sqrt3)=12.928$.
These results clearly indicate the close similarity of the structures we 
obtain to ideal dodecagonal quasicrystals.

So far we have only looked at the structure that results for one particular
set of conditions. Figure \ref{fig:5patch_map} provides an overview of
the structural behaviour as a function of pressure and patch width, and
Figure \ref{fig:5patch_configs} provides example final configurations for different
patch widths at a representative pressure.

In the top right corner of the ($\sigma_{\rm pw}$,$P$) plane, i.e.\ larger patch widths 
and higher pressures, the hexagonal phase is most stable (Figure \ref{fig:xtals}). As one moves away from this corner of the parameter space to
lower pressures and patch widths, the onset of quasicrystal formation is 
signalled by a sharp increase in the number of $\sigma$ environments at the
expense of hexagonal environments (Fig.\ \ref{fig:5patch_map}(a) and (b)).
Close to this boundary, our measure of the twelvefoldness of the diffraction
pattern generally has high values, although this measure should be interpreted
with caution as false positives can arise. For example, the superposition
of diffraction patterns from two different hexagonal domains leads to an
anomalously high value of the twelvefoldness at 
$P=1.7\,\epsilon\sigma_{\rm LJ}^{-2}$ and $\sigma_{\rm pw}=0.63$.
The position of the hexagonal to quasicrystal boundary in Fig.\ \ref{fig:5patch_map} is noticeably to the left of the zero-temperature $\sigma$-hexagonal
boundary in Fig.\ \ref{fig:xtals} because of the entropic stabilization
of the hexagonal phase mentioned earlier.

As one moves away from this boundary, the number of hexagonal atoms slowly 
decreases towards zero and the degree of twelvefoldness generally drops.
Examining the configurations in Fig.\ \ref{fig:5patch_configs} one sees a 
gradual crossover from the quasicrystalline configurations with hexagonal
atoms at the centre of the characteristic dodecagons to a pure
$\sigma$ phase in Fig.\ \ref{fig:5patch_configs}(b). For example, about a half 
of the domains in Fig.\ \ref{fig:5patch_configs}(d) are pure $\sigma$ and
the rest are quasicrystal-like. This crossover is driven by
the hexagonal environments becoming energetically increasingly unfavourable
as the patches become narrower.

At the narrowest patch width we considered, the low-energy crystals are the
distorted versions of the $\sigma$ and $H$ crystals in which three of the 
patches point directly at neighbouring particles (Sect.\ \ref{sect:xtals}).
These two crystal forms are nearly degenerate, and the structure that results
from annealing is a mixture of the two.
Even in their distorted forms these two crystals can form coherent boundaries 
between them, and alternating series of layers of the two crystal forms can be 
seen in Fig.\ \ref{fig:5patch_configs}(a). 

It is also noticeable from Fig.\ \ref{fig:5patch_map} that for higher 
pressures at the narrowest patch width hexagonal crystals again form. 
This change reflects the energetic destabilization of the $\sigma$ phase at 
these patch widths because the particles can no longer form 
five strong interactions. 

\begin{figure}
\includegraphics[width=8.4cm]{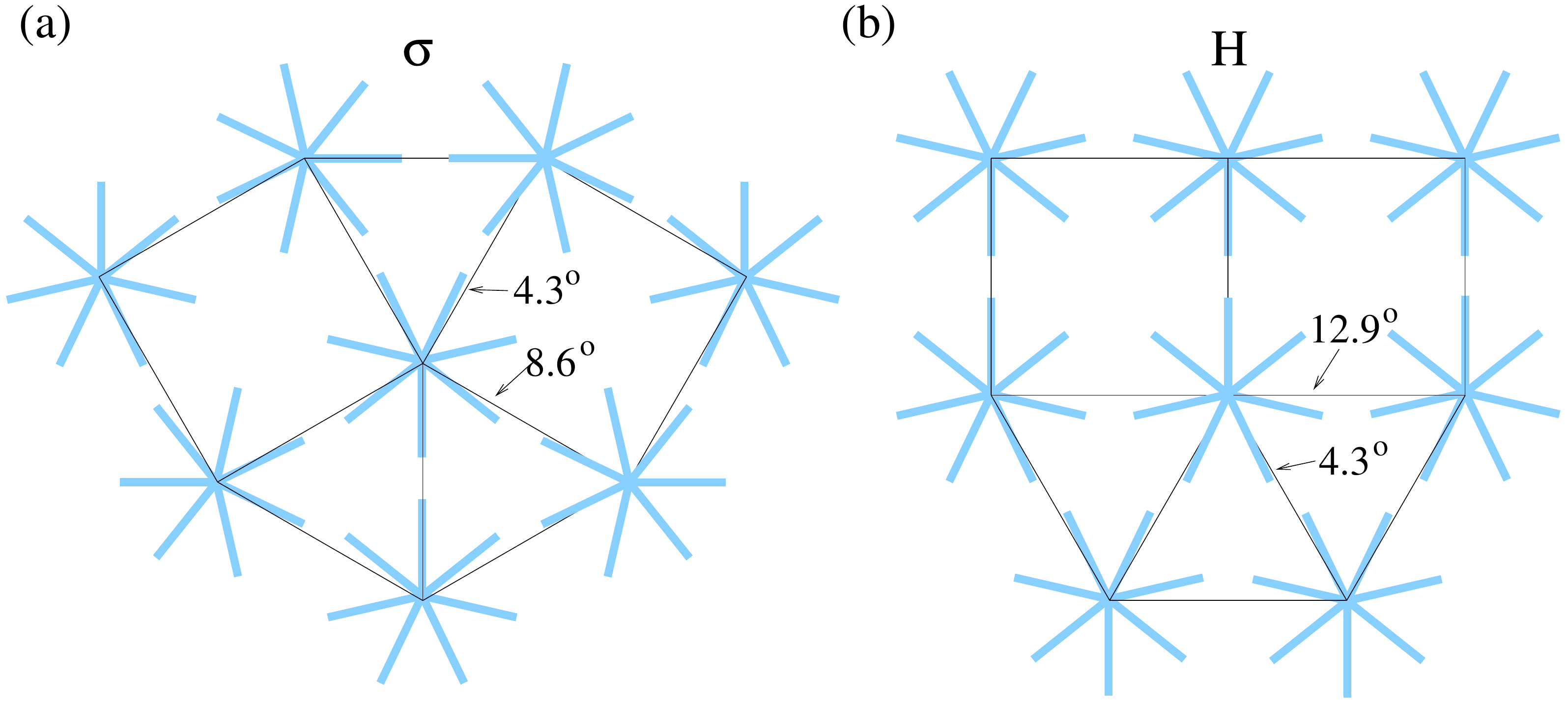}
\caption{\label{fig:7_angles}
Deviations of the patch vectors from the interparticle vectors in the
(a) $\sigma$- and (b) $H$-type environments for 7-patch particles. 
}
\end{figure}

\subsection{7-patch particles}

Like the 5-patch particles, the 7-patch particles have a local symmetry that
is incompatible with crystalline order. However, in addition it is also 
physically unfeasible for a particle to have seven neighbouring atoms at
the equilibrium pair separation
without particles overlapping.
The question is then how do the particles manage to maximize their patch-patch
interactions? 
When the patches are reasonably narrow, the solution the system finds is to 
use only five of the seven patches, and
to adopt the same types of square-triangle tilings as for the 5-patch system.
Figure \ref{fig:7_angles} shows the arrangement of the particles in
the $\sigma$ and $H$ environments, and it can be again seen that the 
$\sigma$ environment has a smaller average deviation of the patch vectors from 
the interparticle vectors. 

Because of this tendency to form square-triangle 
tilings, the behaviour of the 7-patch system is very similar to the 5-patch
system, and so we will much more briefly review the behaviour of this system.
The zero temperature phase diagram shows a very similar form with a hexagonal
crystal having lowest enthalpy at high pressure and patch width, and a $\sigma$
crystal at lower pressures and patch widths. The crossover between these
forms occurs at slightly narrower patch widths than for the 5-patch system, 
because the 7 patches makes the potential somewhat closer to the 
isotropic limit.
As for the 5-patch system at very narrow patch widths, 
this structure distorts so that three of the seven patches
point directly at their neighbours, but this time the acute angle in the 
hexagonal units is $51.43^\circ$ not $72^\circ$.

\label{sect:7patch}
\begin{figure*} 
\includegraphics[width=14cm]{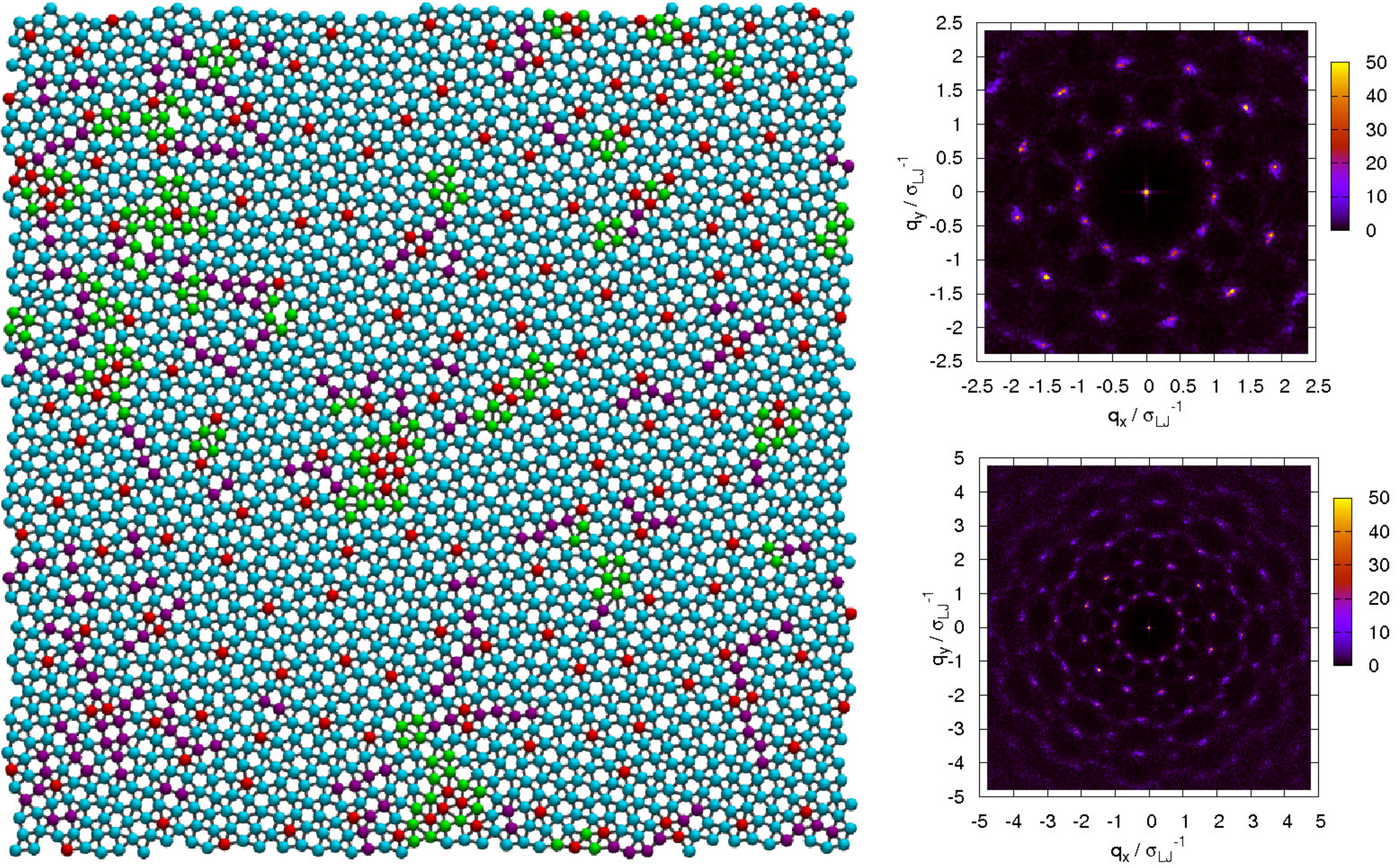} 
\caption{\label{fig:7quasi} 
(a) Configuration after annealing and (b) and (c)
associated diffraction patterns (for two different ranges of $\mathbf{q}$) 
at $p=0.7\,\epsilon\sigma_{\rm LJ}^{-2}$, $\sigma_{\rm pw}=0.42$ for a system of 7-patch particles. 
}
\end{figure*}

Again close to the hexagonal/$\sigma$ boundary our annealing simulations result
in quasicrystalline configurations with clear 12-fold diffraction patterns
(Fig.\ \ref{fig:7quasi}). Inspection of the configurations also again shows
the same dodecagonal motifs (Fig.\ \ref{fig:dodecagon_motifs}) as found for the 
5-patch system. 

\section{Conclusions}

Here we have presented results for a simple two-dimensional patchy particle 
system that exhibits a rich ordering behaviour, in particular showing the 
formation of dodecagonal quasicrystals based on square-triangle tilings for 
certain parameter ranges for particles with five and seven regularly arranged 
patches. 
These examples can be added to the increasing list of model
systems that have been shown to form quasicrystals in simulations.
\cite{Widom87,Leung89,Jagla98,Skibinsky99,Dzugutov93,Quandt99,Keys07,Engel07,Engel10,Iacovella11,HajiAkbari09,HajiAkbari11b,Johnston10b,Johnston11} 
The system also provides a nice model system that illustrates how the local
structural propensities can lead to complex globally-ordered structures. 
In particular, quasicrystals were observed in the region of parameter space 
near to where the enthalpically preferred structure changes from 
five-fold to six-fold coordination, because of the presence of both 
coordination environments in the quasicrystal. Interestingly, the structure
of the quasicrystals is quite similar to model square-triangle dodecagonal 
quasicrystals produced by inflation 
rules.\cite{Stampfli88,Oxborrow93,Hermisson97,Zeng06}

One interesting question concerning the quasicrystals that we observe is 
whether they are thermodynamically stable or just a kinetic product. 
As the enthalpy difference between the quasicrystal and the crystal is not 
that large, it is not unreasonable that the quasicrystal could become 
thermodynamically stable due to the entropy associated with the many possible
configurations for the quasicrystal. However, to confirm this hypothesis
would require the free energy of the quasicrystal (or the 
free energy difference between it and the crystal) to be computed, but it
is not clear how this could be achieved. During some of the cooling simulations,
the system passes from a liquid to a hexagonal crystal and then to a 
quasicrystal as the temperature is decreased, showing that there is a parameter
range where the quasicrystal is more stable than the hexagonal crystal.
However, on heating the $\sigma$ crystal, we have never seen it
transform into a quasicrystal, but instead it directly melts.

An interesting contrast to the behaviour we see here is provided
by particles that have five-fold symmetry in their repulsive 
interactions,\cite{Schilling05,Zhao09} rather than their attractions.
Although hard pentagons show an interesting set of solid phases 
structures,\cite{Schilling05} none of them are quasicrystalline.

Another important question is whether a system that shows behaviour similar
to our model could be experimentally realized. Methods to synthesize
patchy colloids and nanoparticles are developing rapidly, but
as far as we are aware, there are none yet available that could produce, say,
particles with a five-fold symmetric distribution of patches. 
By contrast, effectively restricting such colloidal 
systems to two-dimensions would be relatively straightforward. For example,
if there is a density mismatch between the colloids and the solvent,
sedimentation can lead to monolayer formation at the base of the sample, 
as has been done in recent experiments where a two-dimensional Kagome 
lattice was assembled from `triblock Janus' particles.\cite{Chen11b}

Another possible system in which these dodecagonal quasicrystals
could be potentially realized is the multi-arm DNA motifs produced by the
group of Chengde Mao.\cite{He05,He06b,Yan03,He05b,He06,Zhang08,He08,He07} 
Each arm is made of two-parallel double helices
and both have dangling single-stranded ends that allow them to bind to other
such motifs with a well-defined relative orientation. These effective 
torsional constraints on the interactions lead to quasi-two-dimensional
growth be it into sheets\cite{He05,He06b,Yan03,He05b,He06,Zhang08} 
or closed shells.\cite{Zhang08,He08}
Interestingly, similar to the equivalent patchy particles, the 3-, 4- and
6-arm motifs form two-dimensional honeycomb,\cite{He05,He06b} 
square\cite{Yan03,He05b} and hexagonal\cite{He06} lattices. 
Furthermore, the five-arm motifs can form a $\sigma$-phase like lattice.\cite{Zhang08}

However, there are also a number of significant differences between the 
DNA multi-arm motifs and patchy particle systems. Firstly, the central loop 
to which the stiff double helices are connected has a certain degree of 
flexibility allowing the relative angles of the arms to vary. 
By contrast, the positions of the patches in our model are fixed. 

Secondly, the ``valence'' of the DNA motifs is fixed; i.e.\ the five-arm motifs can only ever bond to five other such motifs. By contrast, for our 
patchy particles at intermediate values of the patch width, the particles can 
adopt five-fold or six-fold coordination environments; it is this flexibility
that allows the system to form the quasicrystals. Therefore, if the DNA motifs
are to be able to form a quasicrystal, a mixture of 5- and 6-arm motifs
with the right stoichiometry would be required. For an ideal dodecagonal 
quasicrystal, the required ratio of 5- and 6-arm tiles would be 
$12.928:1$.
However, even with such a mixture, it might be that the system prefers to phase
separate into $\sigma$ and hexagonal crystals, similar to what has
been seen for mixtures of 3- and 4-arm tiles.\cite{He07}

In future work, we tend to explore this possibility further, first by using a 
coarse-grained model of DNA that we have recently 
developed\cite{Ouldridge10,Ouldridge11} to characterize these DNA 
multi-arm motifs, 
including their flexibility and the angular specificity of their interactions. 
Secondly, this information will then be used to create a patchy-particle
representation, where the patch positions are not rigidly fixed but are 
constrained by an internal potential. 

\begin{acknowledgments}
The authors are grateful for financial support from the EPSRC and the Royal Society.
\end{acknowledgments}

\section{Supplementary Information}

\subsection{Model quasicrystal}

As a comparison to the diffraction patterns we have calculated from our 
simulated configurations, here we report the diffraction pattern from
a model dodecagonal quasicrystal. 
Dodecagonal tilings can be created by projection from a four-dimensional space 
but usually include rhombs as well as squares and triangles.
An alternative approach to create dodecagonal square-triangle tilings is to 
use an inflation approach where an increasingly large structure is iteratively
built up by applying rules for scaling simpler motifs (in our case 
squares and triangles) and then replacing them by more complex 
ones.\cite{Stampfli88,Oxborrow93,Hermisson97,Zeng06}
Here we use what has been termed the ``extended Schlottmann'' inflation 
rules.\cite{Hermisson97,Zeng06} The effect of these rules is at each 
iteration to replace each vertex in the square-triangle tiling by one of the 
dodecagons in Fig.\ \ref{fig:dodecagon_motifs}(a) and (b) depending upon the 
local geometry of the vertex.
Supplementary Figure 1(a) 
shows the structure generated after 3 iterations
of these rules starting from an initial centred hexagon. The corresponding
diffraction pattern has clear 12-fold symmetry 
(Supplementary Fig.\ 1(b)) 
and a very similar pattern of peaks as those in Fig.\ \ref{fig:5patch_QC} and 
\ref{fig:7quasi}.

\begin{figure*}
\setcounter{figure}{0}
\includegraphics[width=14cm]{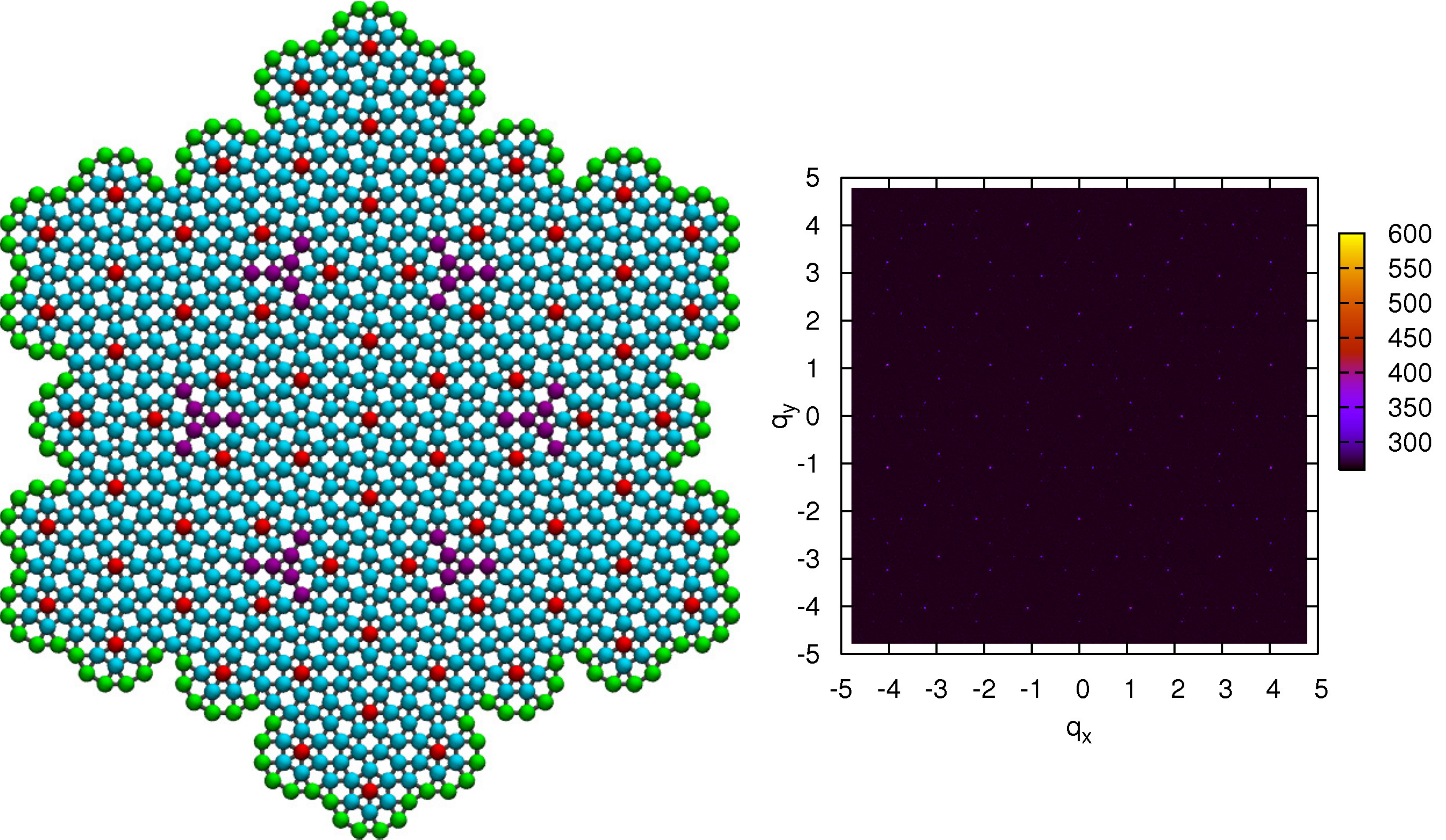}
\caption{\label{fig:perfect} (a) 1063-particle model quasicrystal generated 
by three applications of the extended Schlottmann inflation rules
\cite{Hermisson97,Zeng06} and (b) the associated diffraction pattern.
The resolution we used in (b) was 
$\Delta q_{x}=\Delta q_{y}=0.05\,\sigma_{\rm LJ}^{-1}$.
}
\end{figure*}

\end{document}